\documentclass{article}
\usepackage{fancyvrb}
\usepackage{geometry}
\usepackage{float}
\usepackage{amsmath}
\usepackage{siunitx} 
\usepackage{listings}
\usepackage{graphicx} 
\usepackage{enumerate}
\usepackage{float} 
\usepackage{paralist} 
\usepackage{seqsplit}
\usepackage{array}
\usepackage{multirow}   
\usepackage{hyperref}  
\newcolumntype{L}[1]{>{\centering\let\newline\\\arraybackslash\hspace{0pt}}m{#1}}

\title{\textsc{On the Efficient Design of LSM Stores}} 

\author{MARTIN \textsc{WEISE}\\Vienna University of Technology} 

\date{\today} 
\begin{document}
	
\maketitle 

\begin{abstract}
	
	In the last decade, key-value data storage systems have gained significantly more interest from academia and industry. These systems face numerous challenges concerning storage space- and read optimization. There exists a large potential for improving current solutions by introducing new management techniques and algorithms.
	
	In this paper we give an overview of the basic concept of key-value data storage systems and provide an explanation for bottlenecks. Further we introduce two new memory management algorithms and a improved index structure. Finally, these solutions are compared to each other and discussed.
	
\end{abstract}

\section{Introduction}

Key-value stores have become more popular amongst developers, researchers and companies since the number of applications for which they are used is ever-increasing. In this context, the underlying technology for key-value stores has emerged to be the \textit{Log-structured merge-tree} (LSM-tree). With the principle of LSM-trees, data stores can replace expensive random disk I/O with storage-friendly sequential disk I/O.

Nonetheless, all of this comes with the trade-in of inefficiencies that are not obvious immediately upon looking. Operations that are intended to free up space (compaction) for example are twofold: frequent compactions that enables efficient reads come at a price of slower data store operations, high write amplifications and increased disk wear-out. Since these compactions are part of the LSM-tree principle, they cannot simply be eradicated from the data store. With this in mind, new algorithms need to be developed to counteract these inefficiencies.

This pure literature study mounts primarily on the insights gained from SlimDB \cite{ren2017slimdb}  and Accordion \cite{bortnikov2018accordion}. Additionally, this paper aims to give an overview of key-value data store problems regarding inefficient memory management and high read tail latency, which are the main bottlenecks in key-value data stores that follow the LSM-tree principle (so called \textit{LSM stores}).

The rest of this paper is organized as follows: section \ref{ch:background} gives the reader information about the general LSM store concepts, section \ref{ch:slimdb} introduces the Stepped-Merge memory management algorithm with a new block index structure and section \ref{ch:accordion} proposes an improvement of an LSM-tree implementation. Section \ref{ch:comparison} discusses the aforementioned solutions and finally, section \ref{ch:conclusion} concludes on the learned concepts.

\section{Background}\label{ch:background}

This section addresses the fundamental concepts of memory management in LSM-trees and takes a look at a variant that is used today in a multitude of LSM-tree implementations. Additionally, this paper exhibits performance bottlenecks in established LSM stores. We define LSM-stores to consist of an in-memory storage with a large persistent storage volume that holds a collection of chunks (see Figure \ref{fig:lsmtree}). To better comprehend the new concepts and algorithms presented in this paper, we first describe the basic principle of LSM-trees in detail and their boundaries, as well as the properties they fulfill. Later on, we present implementations of LSM-trees into currently used LSM stores and their limitations as a motivation to argue the need for new memory managements concepts and algorithms.

\subsection{LSM-tree}\label{sec:lsmtree}

With \textit{Log-structured merge-trees} (LSM-trees) write-intensive workloads are first aggregated in a dynamic segment before being flushed onto log-structured persistent storage that is compacted in a background process. Adding ($\mathtt{PUT}$) a new entry to the LSM-tree means first inserting it to the in-memory buffer (and also append to the persistent storage log for crash recovery). Once the buffer is full, a new in-memory buffer is created and the old one is treated as immutable snapshot and flushed to the persistent storage in the background as chunks of data (most implementations name their data structure name, we independently call them \textit{chunks}). Each flush creates a new immutable snapshot and clears the active buffer.

These chunks written to the storage are immutable and allow a fast write to the persistent storage, but result in a notable fraction of old data that is either``overwritten'' (new data is present) or deleted although older copies still exist and thus a waste of persistent storage. Also a search may consists of reading multiple immutable chunks to finally result in the desired value for a given key.

\begin{figure}
	\centering
	\includegraphics[width=.95\textwidth]{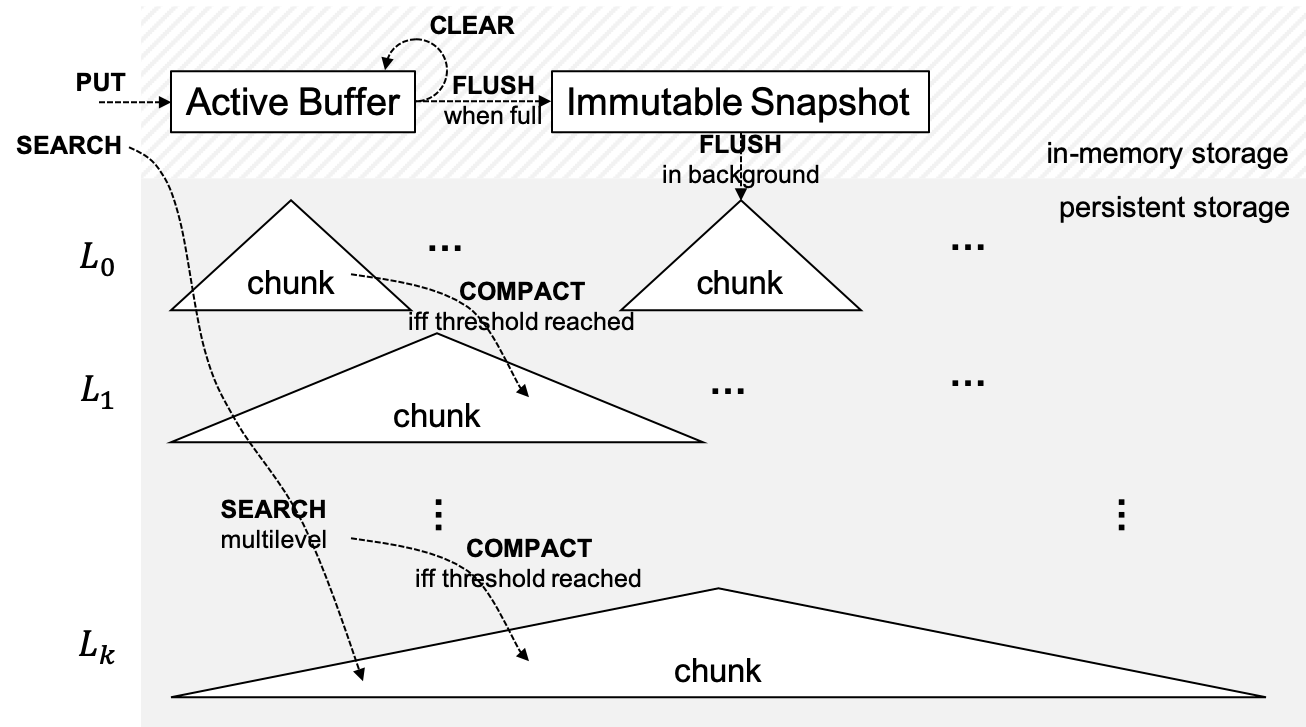}
	\caption{Level structure visualization of a LSM-store.}
	\label{fig:lsmtree}
\end{figure}

To organize the available storage memory efficiently, the compaction process (see Figure \ref{fig:lsmtree}) removes that old data and duplicates that may occurred in multiple chunks before. This compaction process is executed in the background and creates a level-type hierarchy that migrates chunks over time to a deeper level by merging them. The hierarchy follows an exponential growth pattern: each level $i$ is $r$-times larger than the level before $i-1$. As \cite{ren2017slimdb} addresses, common values are $r\in [8,16]$, additionally the maximum number of levels is therefore $\mathcal{O}(\log_r n)$ with $n$ being the number of unique keys. This is also the worst-case lookup time (accessing all levels).

The merging procedure keeps the number of files in a LSM-tree bounded. Again, the paper \cite{ren2017slimdb} addresses the worst-case write amplification to be $\mathcal{O}(r\log_r n)$. The amortized I/O cost per insertion is $\mathcal{O}(r\log_r n)$.

Therefore the operations on a LSM-tree are quite efficient on large $n$ in the context of computational complexity. The sequential access pattern of LSM-trees for writes is the main advantage over other common indexes such as B-trees. Accessing data in a B-tree may involves many random reads/writes, because of their design which makes them not efficient for implementation in storage systems.

\subsection{Bottlenecks}

LSM-trees nowadays are well-studied and fine-tuned. However, with increasing demands for real-time performance and scaling, they face numerous challenges. In this section, we take a look at performance bottlenecks of LSM-trees.

Developers tune the performance of their LSM-stores (implementation of LSM-trees as a data store) with certain performance parameters. This paper should give an overview of general parameters independently of the underlying implementation in LSM-stores.

The first parameter is the rate at which compactions occur. Infrequent compactions makes the store inefficient because multiple versions are scattered across chunks, therefore numerous chunks need to be searched (this also makes caching not economical). Frequent compactions reduce both the space needed to store chunks and their total number. Anyhow, this comes at a much higher CPU and I/O consumption which slows down read/write performance of the store (also caches are invalidated at a higher rate). As \cite{bortnikov2018accordion} points out, the higher write volume accelerates disk wear-out.

However, performance tuning with these parameters only deals with the consequences of organizing persistent storage, not memory management itself. This paper addresses these problems that received much less attention in the sections \ref{ch:slimdb} and \ref{ch:accordion}.

\section{Stepped-Merge}\label{ch:slimdb}

In this section we present the memory management algorithm \textit{Stepped-Merge} and the \textit{Three-level Block Index} from \cite{ren2017slimdb}. To give the reader a solid understanding of the concepts used and how they work together, we first give an overview in section \ref{sec:steppedmerge_overview}, then we present
the algorithm in detail in section \ref{sec:steppedmerge_algorithm}.

\subsection{Overview}\label{sec:steppedmerge_overview}

The LSM-tree has performance tuning parameters that allow developers to optimize the operations based on their application domain. As excpected, the performance gain is somehow always limited to the core design decisions made when initially designing the LSM-tree. With this motivation the Stepped-Merge algorithm alters the memory management and compaction strategy.

Traditional LSM-trees compact entries from a level $i$ with the next level $i+1$. As a consequence, the LSM-tree must at least merge-sort one entry of level $i$ with overlapping entries in level $i+1$. To overcome this write-overhead, the Stepped-Merge algorithm divides entries in each level into $r$ sub-levels. Again, $r$ is a constant commonly chosen to be between $[8, 16]$. The LSM-tree now $r$-way sorts entries on level $i$ and insert them into level $i+1$ as a new sub-level.

As \cite{ren2017slimdb} argues, this additional step only creates a minor overhead and results in a amortized I/O cost of $\mathcal{O}(\frac{1}{C}\log_r n)$ with entries of size $C$ for the $\mathtt{PUT}$ operation. This improves the amortized I/O cost of general LSM-trees and therefore the overall write-amplification of LSM-trees.

\subsection{Algorithm}\label{sec:steppedmerge_algorithm}

This improvement comes at the expense of a increased $\mathtt{SEARCH}$ operation cost with the Stepped-Merge algorithm. The block based indexing mechanism replaces the original LSM- tree array based indexing mechanism entirely. Because the LSM store organization has overlapping sub-levels at each level, the read-performance suffers and involves $\mathcal{O}(r\log_r n)$ random reads. This trade off is unwanted and the algorithm further enhances LSM stores by increasing the in-memory indexes and filters. Filters are omitted in this literature study, we focus on the in-memory indexing mechanism in combination with the algorithm.

In this context another contribution to efficient LSM stores is the \textit{Three-level Block Index} that gradually generates a compact index of variable length for data blocks. This has some advantages: the average space required to store a key can be as low as $8$ bits and improve persistent storage access, because of higher cache hit rates. Additionally the representation can be replaced without a need for changing the LSM store organization and execution. However, this comes at the cost that keys are no longer totally ordered, but semi-sorted (ordered prefix and suffix). Further, the extra-compaction step requires more CPU cycles.

\begin{figure}[H]
	\centering
	\includegraphics[width=.95\textwidth]{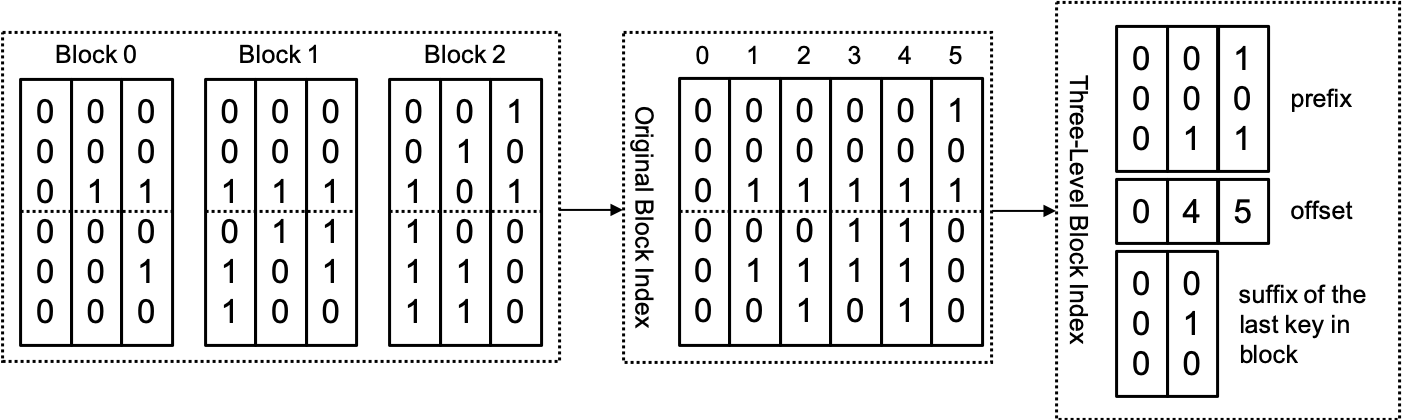}
	\caption{Generation of the Three-level Block Index (image adapted from \cite{ren2017slimdb})}
	\label{fig:tlbi}
\end{figure}

Now we examine the compression-algorithm of the Three-level Block Index which is based on the original LevelDB \cite{ghemawat2019leveldb} implementation of a data block index. It creates an array of unique prefixes and their respective last offset in the original block index (see Figure \ref{fig:tlbi}).

Unfortunately the paper \cite{ren2017slimdb} does not provide (pseudo)code on the creation of a Three-level Block Index. Also, the textual description of the algorithm is not easy to understand for novices. Independently from that, we did an approach to translate the textual description into pseudocode in Listing \ref{lst:tlbi_create}.

rom a cost perspective, the Three-level Block Index consumes at average $10$ bits per block using ECT for the suffix-array. It’s main bottleneck is the need for creating a Entropy-Encoded Trie that consumes CPU cycles. With $16$ entries (key-value pairs) assumed per block this results in a memory overhead of $0.7$ bits per key compared to $8$ using LevelDB. Replacing the original LSM algorithm with Stepped-Merge does not induce additional memory cost, but increased positive-/negative read cost. Stepped-Merge’s main advantage over the traditional LSM algorithm is an improved write cost of $\frac{w}{8}$ bits/key instead of $\frac{rw}{8}$ bits/key (to the best of our knowledge we were not able to examine what $w$ means: it is assumed to denote a write process measured in bits).\\

{\footnotesize
	\begin{lstlisting}[frame=single, caption={Generation process of the Three-level Block Index (Pseudocode)}, label={lst:tlbi_create}, captionpos=b]
Input: Chunk containing data blocks
Output: Three-level Block Index corresponding to the chunk

function generate(chunk) {
  vanilla_block_index = {}
  foreach block in chunk {
    append(vanilla_block_index, block.keys[0])
    append(vanilla_block_index, block.keys[block.keys.length-1])
  }
  prefix_array = {}
  last_offset = {}
  suffixes = {}
  foreach index, key in vanilla_block_index {
    if (changed(key.prefix)) {
      append(prefix_array, key.prefix)
      append(last_offset, index)
    }
  }
  // create ECT from vanilla_block_index
  foreach block in chunk {
    last_key = block.keys[block.keys.length-1]
    if (is_ect_shortest_unique_prefix_path(last_key)) {
      append(suffixes, last_key)
    }
  }
}
	\end{lstlisting}
}

The combination of Stepped-Merge algorithm and Three-level Block Index has similar results compared to the LSM-tree with Three-level Block Index. On average, both consume $2$ bits/key memory and for writes Stepped-Merge with Three-level Block index supersedes the LSM-tree with Three-level Block Index because Stepped-Merge works with blind reads.\\

{\footnotesize
	\begin{lstlisting}[frame=single, caption={$\mathtt{FIND}$ operation in Three-level Block Index (Pseudocode)}, label={lst:tlbi_search}, captionpos=b]
Input: Key of the desired tuple, Full data chunk containing blocks
Output: Value that corresponds to the key

function find(key, block_index) {
  prev = nil
  cur = nil
  foreach index, prefix in block_index.prefix_array {
    if (prefix = key.prefix) {
      cur = index
      break
    }
    prev = index
  }
  // build the ECT trie if not cached
  s = {}
  for suffix in block_index.suffixes {
    if (suffix > key.prefix) {
      s = suffix
    }
  }
  for (; prev < cur; prev++) {
    last_key = block.keys[block.keys.length-1]
    if (last_key.suffix = s) {
      return block[last_key]
    }
    if (last_key.suffix > s) {
      continue
    }
    if (last_key.suffix < s) {
      // get value from this block and return it
    }
  }
}
	\end{lstlisting}
}


\section{Accordion}\label{ch:accordion}

In this section we discuss the memory management algorithm Accordion \cite{bortnikov2018accordion}. To get a good understanding on how the algorithm works, we give a short overview on the algorithms general working principle in section \ref{sec:acc_overview}. The algorithm is then analyzed and explained in detail in section \ref{sec:acc_algorithm}.

\subsection{Overview}\label{sec:acc_overview}

Traditional LSM stores are split into the in-memory storage part and the persistent storage part that is usually designed to be organized in compliance to the LSM design principles. Accordion is a memory store management algorithm that re-applies the principles to the in-memory storage. It resolves the discrepancy of compaction rate tuning by introducing proactive in-memory compactions that delay persistent memory flushes (thus reducing write workload on the persistent storage and disk wear).

An increased number of data inside the in-memory storage part also means that read latency can be improved, because the $\mathtt{SEARCH}$ operation is able to scan more keys in- memory. Additionally frequent $\mathtt{PUT}$ operations do not necessarily trigger compaction procedures a lot. The high-throughput experiments of \cite{bortnikov2018accordion} show that Accordion improves write throughput by up to 48\% while reducing the read tail latency by up to $40\%$ for Zipfian (heavy-tailed) key access distribution. Additionally, the write volume is reduced by up to $30\%$.

\subsection{Algorithm}\label{sec:acc_algorithm}

The algorithm separates the in-memory storage into two units that work independently of each other: \begin{inparaenum}[(1)]
	\item one mutable, small segment that absorbs $\mathtt{PUT}$ operations, and
	\item a sequence of immutable segments that once were a mutable.
\end{inparaenum}

Inside the in-memory storage, it manages indexed data cells as a pipeline of segments where the most recent segment is mutable with $\mathtt{PUT}$ operations. Except for this \textit{active} segment every other segment is immutable. Fetch operations like $\mathtt{SEARCH}$ scan all segments (possible in parallel) inside the in-memory storage before continuing on the persistent storage, if no candidate is found (see Listing \ref{lst:fetching}).\\

{\footnotesize
	\begin{lstlisting}[frame=single, caption={Fetching data in Accordion (Pseudocode)}, label={lst:fetching}, captionpos=b]
Input: Key of desired entry
Output: Value that corresponds to the key

function search(key) {
  foreach chunk in in_memory_storage {
    if (chunk contains key) {
      return chunk.get(key)
    }
  }
  // continue with persistent storage
}
	\end{lstlisting}
}

The active segment size $A$ and the maximum length of the pipeline $S$ can be tuned, while a value of $A\in [0.02,0.05]$ and $S\in [2,5]$ is recommended. Once the threshold $A$ is reached, the active buffer becomes immutable. Listing \ref{lst:put} shows the complete process of adding a key-value tuple to the LSM store.\\

{\footnotesize
	\begin{lstlisting}[frame=single, caption={$\mathtt{PUT}$ operation in Accordion (Pseudocode)}, label={lst:put}, captionpos=b]
Input: Key-value tuple, Active segment threshold A
Output: {}

function put(key, value) {
  segment = active_segment
  // write key-value tuple to active segment
  if (size(active_segment) > A) {
    active_segment.mutable = false
    segment = active_segment
    active_segment = {}
  }
  if (length(pipeline) > S) {
    // merge flat segments in pipeline (ignore duplicates)
    // perform compaction on flattened segment
    // add flat segment to snapshot set (in-memory)
  } else {
    // flatten segment (e.g. replace dynamic skiplist with ordered array)
    // add flattened segment to pipeline
  }
  if (full(in_memory_store)) {
    // normal LSM flush using snapshot technique
  }
}
	\end{lstlisting}
}

\begin{figure}[H]
	\centering
	\includegraphics[width=\textwidth]{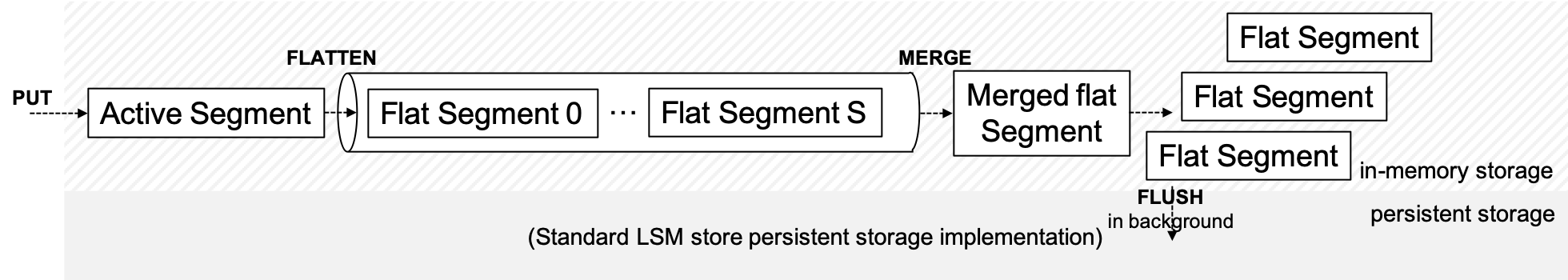}
	\caption{Visualization of the Accordion architecture (image adapted from \cite{bortnikov2018accordion})}
	\label{fig:acc_architecture}
\end{figure}

The $\mathtt{FLATTEN}$ operation in Figure \ref{fig:acc_architecture} informally just replaces dynamic structures (e.g. skiplists) with ordered arrays, the data cells are not re-organized. This optimizes the memory footprint and thus further delays persistent storage flushes and also improves read-latency because more segments are stored in the in-memory storage. Once the pipeline exceeds it’s maximum number of entries inside $S$, it performs a $\mathtt{MERGE}$ of the segments in the pipeline by eliminating redundant data. The resulting flat segments are now snapshots and are $\mathtt{FLUSH}$-ed to the LSM store in the traditional way as already described in section \ref{sec:lsmtree}.

This elimination of redundant data during the $\mathtt{MERGE}$ operation comes with a overhead of additional I/O and CPU cycles. As \cite{bortnikov2018accordion} notes, this is particularly important in heavy-tailed (e.g. Zipfian) distributions that regularly overwrite some keys. This compaction is defined in the form of three policies:

\begin{inparaenum}[(1)]
	\item \textit{Basic}, some redundant elimination is done;
	\item \textit{Eager}, immediate redundant elimination is done; and 
	\item \textit{Adaptive}, that is a mix of Basic and Eager and uses a heuristic based on the level of redundancy.
\end{inparaenum}

Some extensive experiments with all three compaction policies were executed by \cite{bortnikov2018accordion}. Since the whole discussion would push the boundaries of this course, some results are omitted in detail here. However, their results include that the Basic compaction policy improves throughput on SSDs by $47.2\%$ and $25.4\%$ for Zipfian (heavy-tail) distributed keys. All reported results are summarized in Table \ref{tbl:acc_results}.

\begin{table}[H]
	\centering
	\begin{tabular}{|c|c||c|c|}
		\hline
		Compaction & Distribution & SSD & HDD\\
		\hline
		\hline
		\multirow{2}{*}{Basic ($A=0.02, S=5$)} & Zipf & $47.6\%$ & $25.4\%$\\
		\cline{2-4}
		& Uniform & $23.8\%$ & $8.9\%$\\
		\hline
		\multirow{2}{*}{Eager ($A=0.25, S=2$)} & Zipf & \multicolumn{2}{c|}{\multirow{2}{*}{Below Baseline}}\\
		\cline{2-2}
		& Uniform & \multicolumn{2}{c|}{~} \\
		\hline
		Adaptive 					& Zipf & $45.3\%$ & $24.4\%$\\
		\hline
	\end{tabular}
	\caption{Improvements of write throughput based on the experimental results of different compaction policies and distributions as reported in \cite{bortnikov2018accordion}}
	\label{tbl:acc_results}
\end{table}

\section{Comparison}\label{ch:comparison}

In this section we discuss similarities and differences between the two memory management algorithms, Stepped-Merge in section \ref{ch:slimdb} and Accordion in section \ref{ch:accordion}. First we take a short overview on both solutions in section \ref{sec:overview}  and discuss the advantages and drawbacks of both in section \ref{sec:discussion}.

\subsection{Overview}\label{sec:overview}

The presented solutions Accordion and Stepped Merge both address the same problems with LSM-trees: overhead through compaction, no optimization for SSDs (low read performance) and a fragmented memory layout. The algorithms deal with these major bottlenecks in a similar way and aim for the same goal: improving write throughput while reducing overhead caused by compaction. Accordion proposes to re-apply the LSM design principles to the in-memory storage of the LSM-tree to speed-up write operations and increase cache hits. Stepped-Merge replaces the standard compaction procedure of LSM-trees by introducing sub-levels that improve the amortized I/O cost of writes.

\subsection{Discussion}\label{sec:discussion}

Both solutions increase the performance of traditional LSM stores. Accordion introduced three policies to tune the compaction rate. Experiments showed that the Adaptive policy is the best strategy for Accordion by flushing the in-memory store based on a heuristic (more redundancy equals less cost-effectiveness upon compaction). With their recommended parameters they were able to improve write loads by $23.8\%$ to $47.6\%$ for SSD persistent storage and $8.9\%$ to $25.4\%$ for HDD persistent storage. The paper \cite{bortnikov2018accordion} did extensive experiments with different workloads and reported improvements for all of them. Surprisingly, their experiments showed that in many different scenarios, disk I/O is not the principal bottleneck.

The Stepped-Merge algorithm replaced array based indexing mechanisms in LSM stores with a block based indexing mechanism (Three-level Block Index). The key compression technique further allows a large reduction in required storage space for each key, increasing the number of cache hits and the time a key resides in-memory. This somehow comes at the cost of more expensive scanning operations, as more keys have to be searched and duplicates are not eliminated a long time. The paper evaluated the possible combinations of the Stepped-Merge algorithm together with LSM-trees and concluded that all of them improve the original LSM-tree.

\section{Conclusion}\label{ch:conclusion}

With this literature study the principal bottlenecks of LSM stores are addressed at a reasonable level of detail for novices and the need for improvement of existing concepts is discussed. We presented two algorithms that greatly reduce these bottlenecks at a moderate trade-off cost.

LSM-trees are present for a long time and only until a few years ago, researchers picked up the topic and started changing the memory management, because parameter tuning did not deliver the desired results. Future work can pick up from here and further improve LSM-trees based on the findings of Accordion and Stepped-Merge.

\bibliographystyle{ieeetr}
\bibliography{intro}

\begin{thebibliography}{1}

\bibitem{ren2017slimdb}
K.~Ren, Q.~Zheng, J.~Arulraj, and G.~Gibson, ``Slim{D}{B}: {A}
  {S}pace-efficient {K}ey-value {S}torage {E}ngine for {S}emi-sorted {D}ata,''
  {\em Proceedings of the VLDB Endowment}, vol.~10, pp.~2037--2048, Sept. 2017.

\bibitem{bortnikov2018accordion}
E.~Bortnikov, A.~Braginsky, E.~Hillel, I.~Keidar, and G.~Sheffi, ``Accordion:
  better memory organization for {L}{S}{M} key-value stores,'' {\em Proceedings
  of the VLDB Endowment}, vol.~11, pp.~1863--1875, 08 2018.

\bibitem{ghemawat2019leveldb}
S.~Ghemawat and J.~Dean, ``{LevelDB is a fast key-value storage library written
  at Google that provides an ordered mapping from string keys to string
  values.}.'' \url{https://github.com/google/leveldb}, 2019.
\newblock [Online; accessed 06-November-2019].

\end{thebibliography}
	
\end{document}